\documentclass[a4paper]{article}
\usepackage{lmodern}
\usepackage{microtype}

\usepackage{xcolor}
\definecolor{MyDarkBlue}{rgb}{0.15,0.25,0.45} 
\usepackage[
linktocpage=true,
hypertexnames=false,
colorlinks=true,
citecolor=MyDarkBlue,
linkcolor=MyDarkBlue,
urlcolor=MyDarkBlue,
pdfauthor={
    Simon-Raphael Fischer,
    Mehran Jalali Farahani,
    Hyungrok Kim,
    Christian Saemann},
pdftitle={Topological Classification of Spontaneous Symmetry Breaking and Vacuum Degeneracy},
pdfsubject={hep-th},
breaklinks=true
]{hyperref}

\usepackage{amsmath,amssymb,amsthm,booktabs,pdflscape}
\usepackage{cleveref}
\usepackage{mathtools}
\usepackage{bbm}
\usepackage{bm}
\usepackage{mathrsfs}
\usepackage{tikz}
\usetikzlibrary{matrix,cd,arrows}
\usepackage{ifthen}
\newcommand{\makecommand}[3]{%
    \foreach \i in #3 {%
        \expandafter\xdef\csname #1\i\endcsname{\noexpand#2{\unexpanded\expandafter{\i}}}%
    }%
}
\newcommand{\latinalphabet}{A,a,B,b,C,c,d,D,E,e,F,f,G,g,H,h,I,i,J,j,K,k,L,l,M,m,N,n,O,o,P,p,Q,q,R,r,S,s,T,t,U,u,V,v,W,w,X,x,Y,y,Z,z}
\makecommand{I}{\mathbbm}{\latinalphabet}
\makecommand{bf}{\mathbf}{\latinalphabet}
\makecommand{bm}{\bm}{\latinalphabet}
\makecommand{ca}{\mathcal}{\latinalphabet}
\makecommand{fr}{\mathfrak}{\latinalphabet}
\makecommand{rm}{\mathrm}{\latinalphabet}
\makecommand{sf}{\mathsf}{\latinalphabet}
\makecommand{sf}{\mathsf}{{id}}
\makecommand{sc}{\mathscr}{\latinalphabet}

\newcommand{\parder}[2][]{%
    \ifthenelse{\equal{#1}{}}{%
        \frac{\partial}{\partial #2}%
    }{%
        \frac{\partial #1}{\partial #2}%
    }%
}
\newcommand{\delder}[2][]{%
    \ifthenelse{\equal{#1}{}}{%
        \frac{\delta}{\delta #2}%
    }{%
        \frac{\delta #1}{\delta #2}%
    }%
}

\def\spiral[#1][#2][#3:#4:#5]
{
\pgfmathsetmacro{\domain}{pi*#3/180+#4*2*pi}
\draw [#1,shift={(#2)}, domain=0:\domain,variable=\t,smooth,samples=int(\domain/0.08)] plot ({\t r}: {#5*\t/\domain})
}

\begin{document}
    \title{Topological Classification of Symmetry Breaking and Vacuum Degeneracy}
    \author{Simon-Raphael Fischer\footnote{National Center for Theoretical Sciences, Mathematics Division, National Taiwan University, Taipei 106319, Taiwan}\;\footnote{Mathematisches Institut, Georg-August-Universität Göttingen, Bunsenstrasse 3–5, 37073, Göttingen, Germany}\and Mehran Jalali Farahani\footnote{Maxwell Institute for Mathematical Sciences, Department of Mathematics, Heriot--Watt University, Edinburgh EH14 4AS, United Kingdom}\;\footnote{School of Mathematics and Physics, University of Surrey, Guildford GU2 7XH, United Kingdom}\and
    Hyungrok Kim\footnote{Department of Physics, Astronomy and Mathematics, University of Hertfordshire, Hatfield,
Herts.\ AL10 9AB, United Kingdom}\and Christian Saemann\footnotemark[3]
}
\date{
    \texttt{simon.fischer@mathematik.uni-goettingen.de},\quad\texttt{m.jalalifarahani@surrey.ac.uk},\\\texttt{h.kim2@herts.ac.uk},\quad\texttt{c.saemann@hw.ac.uk}\\\today}

    \maketitle

    \begin{abstract}
        We argue that a general system of scalar fields and gauge fields manifesting vacuum degeneracy induces a principal groupoid bundle over spacetime and that the pattern of spontaneous symmetry breaking and the Higgs mechanism are encoded by the singular foliation canonically induced on the moduli space of scalar vacuum expectation values by the Lie groupoid structure. Recent mathematical results in the classification of singular foliations then provide a qualitative classification of the possible patterns of vacuum degeneracy.
    \end{abstract}

    \section{Introduction and summary}
    The twin phenomena of spontaneous symmetry breaking and the Higgs mechanism are foundational cornerstones of our current understanding of the universe, ranging from the origin of the masses of fundamental particles to the appearance of superconductivity and superfluidity.
    Both cases are characterized by a degeneracy of vacua, where the space of allowed vacua forms a manifold, the \emph{vacuum moduli space}.
    
    The familiar examples, such as electroweak symmetry breaking, are relatively simple, consisting of a linear sigma model to which a gauge group is coupled linearly. 
    However, this is far from the most general situation in which spontaneous symmetry breaking and the Higgs mechanism can operate. One can imagine scalar fields living in more complicated manifolds, perhaps taking values in a topologically nontrivial fiber bundle, and gauge fields coupling to the scalar field in more complicated ways. What, then, is the general pattern of vacuum degeneracy, spontaneous symmetry breaking and the Higgs mechanism? Can we hope to classify such patterns? Previously such patterns have been analyzed in special (e.g.~supersymmetric) cases via Hasse diagrams \cite{Bourget:2019aer}, but a more general perspective has been lacking.
    
    In this paper, we argue that recent mathematical results in the theory of singular foliations from differential geometry enable us to understand all possible patterns of vacuum degeneracy. Concretely, we explain the following.
    \begin{itemize}
        \item A general physical system with partially gauged scalar fields is given by the data of a \emph{principal groupoid bundle with connection} that generalizes principal bundles (for the gauge field alone) and fiber bundles (for the scalar field alone).
        \item In such a theory, deformations of the scalar vacuum expectation value (VEV) in the moduli space of scalar VEVs can be of two kinds --- S-type or G-type --- depending on whether an experimenter who has access to the boundary (but not the interior) of an enclosed region can affect a deformation of the VEV inside the region. Accessibility by G-type deformations (which are also those that can be absorbed by gauge transformations) defines a singular foliation on the moduli space of scalar VEVs.
    \item Without knowing the detailed physics (e.g.\ details of the gauge groupoid), the mere knowledge of the topology of the vacuum orbit (i.e.~the subspace of vacua accessible by S-deformations) of a point $v$ in the moduli space of scalar VEVs tightly constrains the possible patterns of S- or G-type deformations in the neighborhood of $v$. Given a vacuum orbit topology and a putative transverse model of the transverse deformations, it is possible to effectively compute whether the transverse model is in fact possible or not.
    \end{itemize}
    Overall, we have a dictionary between the physics of vacuum degeneracy and the mathematics of Lie groupoids and singular foliations as given in \cref{table:dictionary}.
    
    \begin{landscape}
    \begin{table}\centering
    \begin{tabular}{ll}\toprule
    \emph{Physics} & \emph{Mathematics} \\\midrule
    Vacuum moduli space of scalar field & Base manifold \(M_0\) of gauge algebroid \\
    IR kinematics of theory with vacuum degeneracy & Gauge algebroid \(E\to M_0\) \\
    Vacuum orbit \(L\) under S-deformations not crossing phase boundary & Leaf \(L\) of singular foliation\\
		\hspace{.5cm} with \(k\) massive vectors and \(d\) Goldstone bosons & \hspace{.5cm} of dim.\ \(k\) and codim.\ \(d\) \\
    Phase transition & Change in dimension of leaves of \(M_0\) \\
    Pattern of vacuum deformations to/from \(L\) & Transverse model of leaf \(L\)\\
    Lie algebra of unbroken remaining gauge symmetry & Isotropy Lie algebra of \(E\) restricted to leaf \(L\) \\
    S-type deformations & Tangent vector of \(M_0\) tangent to leaves\\
    G-type deformations and phase-transitioning S-type deformations & Tangent vector of \(M_0\) transverse to leaves \\
    Higgs bosons or H-type deformations & Tangent vector of \(M\) transverse to \(M_0\) \\
    \bottomrule
    \end{tabular}
    \caption{Dictionary between the mathematics of foliations and the physics of vacuum degeneracy}\label{table:dictionary}
    \end{table}
    \end{landscape}
    
    \section{General ansatz for scalar--gauge systems}
    
    Let us consider a general local field theory containing a scalar field. We do not assume Lorentz symmetry, merely locality, gauge invariance, and homogeneity of physics at different points of a spacetime $X$. Under these assumptions, a general scalar field is given by a section
    \begin{equation}
        \Phi\colon X\to F
    \end{equation}
    of a fiber bundle \(F\) over \(X\) whose fiber (the target space of the scalar field) is \(M\).\footnote{This is a consequence of the homogeneity assumption.} Thus, equipped with suitable metrics \(g_{\mu\nu}\) and \(\gamma_{ij}\) on \(X\) and \(M\) respectively, a general scalar field theory Lagrangian density takes the form\footnote{This action can be defined by covering \(M\) with coordinate patches and defining the action patchwise; different patches are glued together using coordinate transformations as usual.}
    \begin{equation}
        \mathcal L = \frac12g^{\mu\nu}\gamma_{ij}(D_\mu\Phi^i)(D_\nu\Phi^j) + V(\Phi) + \dotsb,
    \end{equation}
    where \(D_\mu\Phi^i\) is the derivative of a section of \(F\) and \(V\colon M\to\mathbb R\) is a potential function bounded below with a characteristic energy scale \(m\); the ``\(\dotsb\)'' may include higher-order derivative interactions.
    
    Similarly, consider a general local field theory on spacetime \(X\) containing a gauge field \(A\) with gauge group \(G\). The kinematical data are given by a connection on a principal bundle \(P\) with structure group \(G\); a general Yang--Mills Lagrangian density takes the form
    \begin{equation}
        \mathcal L = \frac12 F_{a\mu\nu} F^{a\mu\nu} + \dotsb,
    \end{equation}
    where
    \begin{equation}
        F^a_{\mu\nu}=\partial_\mu A^a_\nu-\partial_\nu A^a_\mu+f^a{}_{bc}A^b_\mu A^c_\nu
    \end{equation}
    is the field strength and where ``\(\dotsb\)'' may include higher-order interactions.
    
    When one has both a scalar field and a gauge field, however, the most general structure compatible with homogeneity on \(X\) is a \emph{principal groupoid bundle} \cite[§5.7]{Moerdijk:2003bb}. Instead of a scalar manifold \(M\) and a Lie group \(G\) separately, the two structures combine into a \emph{Lie groupoid}, which we call the gauge groupoid.\footnote{That is, the structural groupoid of the principal groupoid bundle; not to be confused with the Atiyah groupoid in the mathematical literature.} The infinitesimal version of a Lie groupoid, the analogue of the Lie algebra associated to a Lie group, is called a \emph{Lie algebroid}, which is a vector bundle \(E\) over a manifold \(M\) together with a Lie bracket \([s_1,s_2]^a=f^a{}_{bc}s_1^bs_2^c\) between sections of \(E\) and a map (called the \emph{anchor}) \(s^a\mapsto \rho_a^is^a\) from sections of \(E\) to vector fields on \(M\) that satisfy appropriate compatibility conditions (see \cite{Mackenzie:1987aa,0521499283,Moerdijk:2003bb} for details).
    
    The gauge theory associated to a principal groupoid bundle \cite{Strobl:2004im,Kotov:2015nuz,Fischer:2021yoy,Fischer:2021glc,Fischer:2022sus,Fischer:2020lri,Fischer:2024vak} then describes a scalar field \(\Phi\) coupled to a gauge field \(A^a\) according to the Lagrangian density
    \begin{equation}
        \mathcal L = \frac12D_\mu\Phi^a D^\mu\Phi_a + V(\Phi) + \frac12F^a_{\mu\nu}F_a^{\mu\nu}+\dotsb,
    \end{equation}
    where the scalar field's covariant derivative is
    \begin{equation}
        D_\mu\Phi^i = \partial_\mu\Phi^i -\rho_a^i(\Phi)A^a_\mu,
    \end{equation}
    and the gauge boson field strength is
    \begin{multline}
        F^a_{\mu\nu} = \partial_\mu A_\nu^a - \partial_\nu A_\mu^a + f^a{}_{bc}(\Phi) A_\mu^b A_\nu^c + \omega^a_{bi}(\Phi) D_\mu \Phi^i A^b_\nu
				\\
				- \omega^a_{bi}(\Phi) D_\nu \Phi^i A^b_\mu + \zeta^a_{ij}(\Phi)D_\mu\Phi^iD_\nu\Phi^j.
    \end{multline}
    For the kinematics to be gauge-invariant, the coefficients \(\rho_\mu{}^a_b\) and \(\zeta^a_{ij}\) must satisfy the compatibility conditions \cite[Thm.\ 4.7.5]{Fischer:2021glc}
    \begin{subequations}
    \label{eq:adjustment_cond}
    \begin{align}
        \nabla_i(f^a{}_{bc}-2\rho^j_{[b}\omega^a_{c]j})
        &= 2\rho^j_{[b}R_\nabla{}^a_{c]ji},
        \\
        (R_\nabla)^a_{ijb}
        &=-(\mathrm d^{\nabla^\mathrm{bas}}\zeta)^a_{ijb},
    \end{align}
    \end{subequations}
    where \(R_\nabla{}^a_{bij}\) is the curvature associated to the connection \(\omega^a_{bi}\) and \(\mathrm d^{\nabla^\mathrm{bas}}\) is the covariant exterior derivative with respect to the basic connection \(\nabla^\mathrm{bas}\) associated to the connection \(\omega^a_{bi}\) (see \cite{0521499283,Moerdijk:2003bb,Fischer:2021glc} for the relevant definitions).
    In mathematical terms, \(\omega^a_{bi}\) defines a \emph{Cartan connection} on the gauge groupoid such that its curvature has a \emph{primitive} \(\zeta\), an additional structure function that appears in the field strengths. As shown in~\cite{Fischer:2024vak}, this structure corresponds to an adjustment in the sense of higher gauge theory \cite{Sati:2008eg,Sati:2009ic,Fiorenza:2010mh,Saemann:2019dsl,Kim:2019owc,Borsten:2021ljb,Rist:2022hci}.\footnote{Furthermore, the metric \(\gamma_{ij}\) must be invariant, just like the Killing form on a semisimple Lie algebra, and the potential \(V\) and any other interaction terms must be invariant under the gauge symmetry, as usual.}
    
    Now, consider the effective theory in a regime \(E\ll m\) where \(m\) is the characteristic energy scale of the scalar potential \(V\). Then scalar fields become heavy and can be integrated out except for those along the minima of \(V\);
    let \(M_0\subset M\) be the subspace corresponding to the minima of \(V\).\footnote{
        That is, the subset of critical points of \(V\) that are stable (i.e.~local minima) (if we are purely classical) or global minima (if we allow quantum tunneling considerations).
    }
    Thus, the low-energy kinematics is entirely controlled by the restricted Lie algebroid \(E|_{M_0}\).
    
    \section{The ridged landscape of vacuum deformations}
    \paragraph{G-type and S-type vacuum deformations.}
    For a scalar--gauge theory as described above, we have the scalar target space \(M\), inside which lies a continuously parameterized subspace \(M_0\subset M\) of vacua given by the minimum of the potential \(V\), around which we can expand our fields. But this space is not structureless. At a given vacuum \(v\in M_0\) in the vacuum moduli space, we can distinguish between three types of deformations within \(M\) (elements of the tangent space \(\mathrm T_vM\)), of which the first two are vacuum deformations (elements of \(\mathrm T_vM_0\subset\mathrm T_vM\)): G-type (for \emph{Goldstone} or \emph{general}) deformations, S-type (for \emph{Stueckelberg} or \emph{special}) deformations, and H-type (for \emph{Higgs}) deformations. The latter are easily characterized: they are deformations not in \(\mathrm T_vM_0\), directions along which the potential is not flat, and hence correspond to massive modes such as the ones that arise in the Higgs mechanism. On the other hand, the two types (G-type and S-type) of vacuum deformations differ depending on how the deformation relates to the gauge symmetry.
    
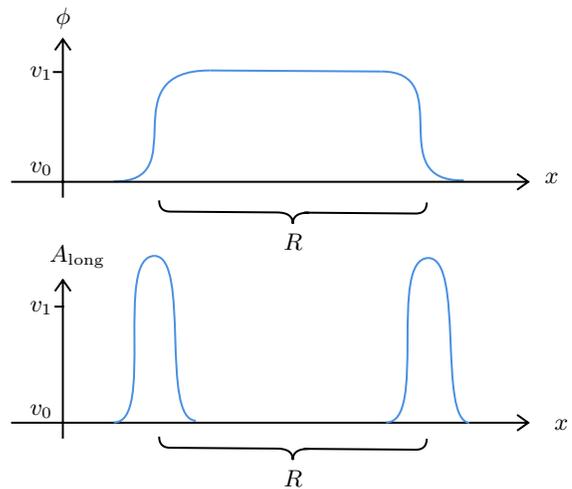
\begin{figure}\centering

\tikzset{every picture/.style={line width=0.75pt}}

\begin{tikzpicture}[x=0.75pt,y=0.75pt,yscale=-0.8,xscale=0.8]
\draw [color={rgb, 255:red, 74; green, 144; blue, 226 }  ,draw opacity=1 ]   (119.67,111.07) .. controls (171.67,111.73) and (116.33,41.07) .. (180.33,41.07) ;
\draw [color={rgb, 255:red, 74; green, 144; blue, 226 }  ,draw opacity=1 ]   (340.33,110.4) .. controls (288.33,111.73) and (337,41.73) .. (288.33,41.73) ;
\draw [color={rgb, 255:red, 74; green, 144; blue, 226 }  ,draw opacity=1 ]   (180.33,41.07) -- (288.33,41.73) ;
\draw  (55,111.33) -- (381,111.33)(87.6,21.33) -- (87.6,121.33) (374,106.33) -- (381,111.33) -- (374,116.33) (82.6,28.33) -- (87.6,21.33) -- (92.6,28.33)  ;
\draw  (55,263.33) -- (381,263.33)(87.6,173.33) -- (87.6,273.33) (374,258.33) -- (381,263.33) -- (374,268.33) (82.6,180.33) -- (87.6,173.33) -- (92.6,180.33)  ;
\draw   (148.33,123.07) .. controls (148.3,127.74) and (150.61,130.09) .. (155.28,130.12) -- (222.61,130.65) .. controls (229.28,130.7) and (232.59,133.06) .. (232.56,137.73) .. controls (232.59,133.06) and (235.94,130.76) .. (242.61,130.81)(239.61,130.78) -- (309.95,131.34) .. controls (314.62,131.37) and (316.97,129.06) .. (317,124.39) ;
\draw    (87.33,42.07) -- (81.33,42.07) ;
\draw [color={rgb, 255:red, 74; green, 144; blue, 226 }  ,draw opacity=1 ]   (119.67,263.07) .. controls (146.13,264.07) and (119.47,158.07) .. (145.47,158.07) ;
\draw    (88.33,190.07) -- (82.33,190.07) ;
\draw [color={rgb, 255:red, 74; green, 144; blue, 226 }  ,draw opacity=1 ]   (291.33,263.07) .. controls (317.8,264.07) and (292.13,159.4) .. (318.13,159.4) ;
\draw [color={rgb, 255:red, 74; green, 144; blue, 226 }  ,draw opacity=1 ]   (171.47,262.07) .. controls (148.8,261.4) and (168.13,158.73) .. (145.47,158.07) ;
\draw [color={rgb, 255:red, 74; green, 144; blue, 226 }  ,draw opacity=1 ]   (343.8,263.4) .. controls (323.8,261.4) and (340.8,160.07) .. (318.13,159.4) ;
\draw   (148.33,272.07) .. controls (148.3,276.74) and (150.61,279.09) .. (155.28,279.12) -- (222.61,279.65) .. controls (229.28,279.7) and (232.59,282.06) .. (232.56,286.73) .. controls (232.59,282.06) and (235.94,279.76) .. (242.61,279.81)(239.61,279.78) -- (309.95,280.34) .. controls (314.62,280.37) and (316.97,278.06) .. (317,273.39) ;
\draw (65,37) node [anchor=north west][inner sep=0.75pt]    {$v_1$};
\draw (65,97) node [anchor=north west][inner sep=0.75pt]    {$v_0$};
\draw (389.67,103.73) node [anchor=north west][inner sep=0.75pt]    {$x$};
\draw (225,142.4) node [anchor=north west][inner sep=0.75pt]    {$R$};
\draw (395.67,259.73) node [anchor=north west][inner sep=0.75pt]    {$x$};
\draw (81.33,-0.27) node [anchor=north west][inner sep=0.75pt]    {$\phi $};
\draw (65,185) node [anchor=north west][inner sep=0.75pt]    {$v_1$};
\draw (65,250) node [anchor=north west][inner sep=0.75pt]    {$v_0$};
\draw (77.33,149.4) node [anchor=north west][inner sep=0.75pt]    {$A_\mathrm{long}$};
\draw (225,291.4) node [anchor=north west][inner sep=0.75pt]    {$R$};

\end{tikzpicture}

\caption{Deformations of the vacuum expectation value for G-type and S-type vacuum deformations. Above: G-type deformations require an experimenter with access to the entire region \(R\) in which the vacuum will be deformed since the expectation value of the Goldstone boson \(\langle\phi\rangle\) differs from \(v_0\) in the entirety of \(R\). Below: to perform an S-type deformation, an experimenter only needs access to the boundary \(\partial R\) of the region, since the expectation of the longitudinal mode of the massive vector boson \(\langle A_\mathrm{long}\rangle\) differs from \(v_0\) only near \(\partial R\).}\label{fig:vacuum-deformation-type}
    \end{figure}

    A thought experiment that distinguishes the two types of vacuum deformations is the following: Given a bounded spatial region \(R\subset\mathbb R^3\), suppose that an experimenter wishes to deform the vacuum inside \(R\); to what subset of \(R\) does the experimenter need access? More formally, starting with a field configuration gauge-equivalent to one in which a scalar field takes the value \(\langle\phi(x)\rangle=v_0\) with all non-scalar fields vanishing, to what regions of spacetime does the experimenter need access so as to end up with a field configuration gauge-equivalent to one in which
    \begin{equation}\label{eq:phi-ansatz}
        \langle\phi(x)\rangle=\begin{cases}
            v_1 & \text{if \(x\in R\)} \\
            v_0 & \text{if \(x\not\in R\)}
        \end{cases}
    \end{equation}
    and all non-scalar fields vanish?\footnote{Or, more realistically, \(\phi(x)\) can be some smooth function approximating the above step function, as shown in \cref{fig:vacuum-deformation-type}.}
    We define a \emph{G-type deformation} as one where the experimenter needs access to the entirety of the region \(R\); an \emph{S-type deformation}, on the other hand, is defined as one where the experimenter only needs access to the boundary \(\partial R\) of the region (see \cref{fig:vacuum-deformation-type}).
    
    An example of a G-type deformation is a real-valued scalar field \(\phi\colon\mathbb R^4\to\mathbb R\) coupled to a Dirac spinor $\Psi$ according to
    \begin{equation}
        \mathcal L = \frac12(\partial\phi)^2+\bar\Psi\mathrm i\gamma^\mu\partial_\mu\Psi+\lambda\bar\Psi\gamma^\mu(\partial_\mu\phi)\Psi+\dotsb.
    \end{equation}
    This theory has a global translation symmetry \(\phi\mapsto\phi+\epsilon\), which is spontaneously broken by a vacuum expectation value \(\langle\phi(x)\rangle=v_0\) of \(\phi\). We write \(\phi=v_0+\phi_\mathrm{G}\) where \(\phi_\mathrm{G}\) is the massless Goldstone boson of this spontaneously broken symmetry. To modify the vacuum to \(v_1\) inside a given region \(R\subset \mathbb R^3\) is the same as creating a condensate of Goldstone bosons at every point inside \(R\) so as to realize \eqref{eq:phi-ansatz}. For this, the experimenter needs access to every point inside \(R\).
    
    An example of an S-type deformation arises in the Stueckelberg mechanism (reviewed in \cite{Ruegg:2003ps}). Consider the action
    \begin{equation}
        \mathcal L=-\frac14(\partial_\mu A_\nu-\partial_\nu A_\mu)(\partial^\mu A^\nu-\partial^\nu A^\mu)+\frac12(\partial_\mu\phi-mA_\mu)(\partial^\mu\phi-mA^\mu),
    \end{equation}
    which has the gauge symmetry
    \begin{equation}
        \binom{\phi(x)}{A_\mu(x)}\mapsto\binom{\phi(x)+\epsilon(x)}{A_\mu(x)+m^{-1}\partial_\mu\epsilon(x)}.
    \end{equation}
    This action describes a propagating massive vector field with mass \(m\).
    Under the gauge symmetry, a field configuration of the form
    \begin{align}\label{eq:stueckelberg-field-configuration-1}
        \phi(x)&=v_0+(v_1-v_0)\chi_R(x)=\begin{cases}
        v_1 & \text{if \(x\in R\)},\\
        v_0 & \text{if \(x\not\in R\)}
    \end{cases}
    \\
    A_\mu(x)&=0,
    \end{align}
    (where \(\chi_R\) is the characteristic function of \(R\), as defined above)
    is gauge-equivalent to a field configuration of the form
    \begin{align}\label{eq:stueckelberg-field-configuration-2}
        \phi(x)&=v_0, &A_\mu(x)&=\frac{v_0-v_1}m\partial_\mu\chi_R(x)=\frac{v_0-v_1}m\delta_{\partial R}n_\mu,
    \end{align}
    where \(\delta_{\partial R}\) is the Dirac delta function supported on the boundary \(\partial R\) and \(n_\mu\) is the unit normal vector to \(R\) pointing inward.\footnote{The delta function in \eqref{eq:stueckelberg-field-configuration-2} is an artifact of the step function in \cref{eq:stueckelberg-field-configuration-1}; if the step function were smeared into a smooth function as shown in \cref{fig:vacuum-deformation-type}, then \eqref{eq:stueckelberg-field-configuration-2} also becomes smooth.} The derivative $\partial\phi$ has become the longitudinal mode \(A_\text{long}\sim m^{-1}\partial\phi\) of the massive vector field \(A\); all that remains of \(\phi\) is the multiplicity of possible vacua corresponding to \(\langle\phi\rangle\).\footnote{A reader skeptical of the reality of the multiplicity of vacua may instead think of the Higgs mechanism, in which the multiplicity of vacua is better known; the Higgs mechanism reduces to the Stueckelberg mechanism when the mass of the Higgs boson is taken to be infinite while the vector boson mass is held finite.}
    In this case, the would-be Goldstone \(\phi\) drops out of the particle spectrum,\footnote{
        This is easy to see in the so-called unitary gauge. Inside any finite region, one can gauge-fix \(\phi=0\); then what remains is the Proca action for a massive vector boson. One cannot gauge-fix \(\phi=0\) everywhere using gauge transformations that decay at infinity, but the degree of freedom count is a local property of the theory and does not depend on this detail.
    } and only the massive longitudinal mode \(A_\text{long}\sim m^{-1}\partial\phi\) remains. An experimenter wishing to change \(\phi\) inside \(R\) then simply needs to create an appropriate condensate of longitudinal modes of \(Z\) at the boundary of \(R\); there is nothing to do in the interior of \(R\). In this regard, S-type deformations are much easier to perform than G-type deformations.
    
    The two types of vacuum deformations can coexist in a single physical system. For example, let \(\phi\colon\mathbb R^4\to M_0\) be a scalar field taking values in a manifold \(M_0\), and suppose that we gauge a subgroup \(\Gamma\) of the isometry group of \(M_0\). Then the tangent vectors along the flow of the isometries in \(\Gamma\) will be S-type, while other tangent vectors will be G-type.
    
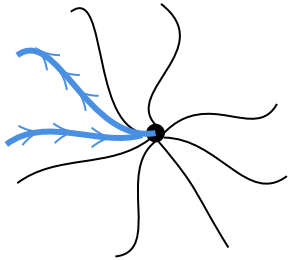
\begin{figure}\centering
\tikzset{every picture/.style={line width=0.75pt}}
\begin{tikzpicture}[x=0.75pt,y=0.75pt,yscale=-0.5,xscale=0.5]
\draw  [fill={rgb, 255:red, 0; green, 0; blue, 0 }  ,fill opacity=1 ] (303,247.55) .. controls (303,242.88) and (306.78,239.1) .. (311.45,239.1) .. controls (316.12,239.1) and (319.9,242.88) .. (319.9,247.55) .. controls (319.9,252.22) and (316.12,256) .. (311.45,256) .. controls (306.78,256) and (303,252.22) .. (303,247.55) -- cycle ;
\draw    (226,125.1) .. controls (266,95.1) and (243.45,258.55) .. (311.45,247.55) ;
\draw    (172,298.1) .. controls (212,268.1) and (265,284.1) .. (305,254.1) ;
\draw    (271,372.1) .. controls (319,367.2) and (271.45,286) .. (311.45,256) ;
\draw    (317,117.1) .. controls (375,162.1) and (279,205.1) .. (311.45,239.1) ;
\draw    (319.9,247.55) .. controls (365,201.1) and (412.1,257.65) .. (434,218.1) ;
\draw    (320,252.1) .. controls (375.1,252.65) and (404,321) .. (444,291) ;
\draw    (314.45,256) .. controls (357,306.1) and (349,305.1) .. (385,363.1) ;
\draw [color={rgb, 255:red, 74; green, 144; blue, 226 }  ,draw opacity=1 ][line width=2.25]    (161,259) .. controls (201,229) and (256,262.1) .. (298,249.55) ;
\draw [color={rgb, 255:red, 74; green, 144; blue, 226 }  ,draw opacity=1 ][line width=2.25]    (172,169.1) .. controls (212,139.1) and (243.45,258.55) .. (311.45,247.55) ;
\draw  [color={rgb, 255:red, 74; green, 144; blue, 226 }  ,draw opacity=1 ] (173.22,240.18) .. controls (180.86,244.31) and (188.04,246.25) .. (194.8,246.02) .. controls (188.48,248.43) and (182.61,253.01) .. (177.18,259.78) ;
\draw  [color={rgb, 255:red, 74; green, 144; blue, 226 }  ,draw opacity=1 ] (208.3,237.75) .. controls (215.36,242.81) and (222.24,245.65) .. (228.97,246.28) .. controls (222.4,247.87) and (216,251.67) .. (209.75,257.7) ;
\draw  [color={rgb, 255:red, 74; green, 144; blue, 226 }  ,draw opacity=1 ] (246.99,242.01) .. controls (253.67,247.56) and (260.33,250.89) .. (267,251.99) .. controls (260.33,253.11) and (253.67,256.44) .. (247.01,262.01) ;
\draw  [color={rgb, 255:red, 74; green, 144; blue, 226 }  ,draw opacity=1 ] (202.08,184.1) .. controls (200.64,175.54) and (197.75,168.67) .. (193.41,163.49) .. controls (199.2,166.99) and (206.42,168.8) .. (215.1,168.92) ;
\draw  [color={rgb, 255:red, 74; green, 144; blue, 226 }  ,draw opacity=1 ] (222.08,204.1) .. controls (220.64,195.54) and (217.75,188.67) .. (213.41,183.49) .. controls (219.2,186.99) and (226.42,188.8) .. (235.1,188.92) ;
\draw  [color={rgb, 255:red, 74; green, 144; blue, 226 }  ,draw opacity=1 ] (241.08,225.1) .. controls (239.64,216.54) and (236.75,209.67) .. (232.41,204.49) .. controls (238.2,207.99) and (245.42,209.8) .. (254.1,209.92) ;
\end{tikzpicture}

    \caption{If one allows S-type vacuum deformations across phase boundaries, the equivalence relation becomes coarse. In the above, we have two phases --- the zero-dimensional, codimension-two leaf (black dot), where gauge symmetry is completely unbroken, and the one-dimensional, codimension-one leaves (black lines), where half of the gauge symmetry is broken. Under the equivalence relation defined by arbitrary S-type deformations that may cross phase boundaries (e.g.~blue curve), all points on the moduli space are equivalent. Under the equivalence relation defined by S-type deformations that do not cross phase boundaries, the equivalence classes correspond to leaves of the singular foliation.}\label{fig:too-coarse}
    \end{figure}

    In the vacuum moduli space \(M_0\), we therefore have an equivalence relation based on whether two points \(\phi_0,\phi_1\in M_0\) can be continuously deformed into each other solely using S-type deformations. A technical point is that, in the presence of distinct phases, that is, different points in the moduli space \(M_0\) where different numbers of Goldstone bosons survive, this equivalence relation becomes too coarse, since points that would otherwise not be connected by S-type deformations may be connected by S-type deformations to and from the symmetry-restoring point, as in \cref{fig:too-coarse}. Thus, it is useful to work with the finer equivalence relation given by S-type deformations that do not cross phase boundaries; the resulting partition is a refinement of the partition of \(M_0\) into phases.
    
    \section{Mathematical description}
\begin{figure}\centering

\tikzset{every picture/.style={line width=0.75pt}}
\begin{tikzpicture}[x=0.75pt,y=0.75pt,yscale=-0.5,xscale=0.5]
\draw  [line width=1] [line join = round][line cap = round] (177,185.5) .. controls (187.81,185.5) and (198.56,191.66) .. (207,197.5) .. controls (210.49,199.92) and (221.62,204.74) .. (224,209.5) .. controls (235,231.51) and (253.35,276.46) .. (276,288.5) .. controls (317.94,310.78) and (350.71,310.45) .. (400,309.5) .. controls (412.07,309.27) and (428.62,310.56) .. (440,305.5) .. controls (450.09,301.01) and (455.72,292.71) .. (464,286.5) .. controls (479.19,275.11) and (497.79,268.37) .. (501,247.5) .. controls (504.12,227.21) and (500.83,204.93) .. (495,185.5) .. controls (490.91,171.86) and (488.81,157.98) .. (482,145.5) .. controls (477.65,137.53) and (469.63,131.07) .. (464,124.5) .. controls (446.23,103.77) and (429.96,89.74) .. (402,84.5) .. controls (392.77,82.77) and (383.33,82.54) .. (374,81.5) .. controls (314.7,74.91) and (247.63,74.66) .. (192,98.5) .. controls (176.69,105.06) and (163.32,116.85) .. (156,131.5) .. controls (154.46,134.58) and (150.32,137.81) .. (150,141.5) .. controls (149.42,148.14) and (149.68,154.84) .. (150,161.5) .. controls (150.35,168.84) and (158.4,174.31) .. (161,179.5) .. controls (161.7,180.91) and (164.59,185.36) .. (166,185.5) .. controls (169.32,185.83) and (172.67,185.5) .. (176,185.5) ;
\draw  [color={rgb, 255:red, 74; green, 144; blue, 226 }  ,draw opacity=1 ][line width=1] [line join = round][line cap = round] (286,186.5) .. controls (323.03,198.84) and (316.43,241.36) .. (331,270.5) .. controls (340.7,289.89) and (375.45,292.35) .. (397,294.5) .. controls (401.74,294.97) and (406.67,297.5) .. (412,297.5) ;
\draw  [color={rgb, 255:red, 74; green, 144; blue, 226 }  ,draw opacity=1 ][line width=1] [line join = round][line cap = round] (291.59,177.54) .. controls (302.33,177.54) and (309.57,186.07) .. (316.21,193.24) .. controls (337.26,215.98) and (340.37,251.25) .. (362.08,271.76) .. controls (381.39,290) and (414.24,288.18) .. (438.15,287.46) .. controls (441.92,287.35) and (451.73,280.21) .. (459.41,280.21) ;
\draw  [color={rgb, 255:red, 74; green, 144; blue, 226 }  ,draw opacity=1 ][line width=1] [line join = round][line cap = round] (320.77,184.63) .. controls (359.4,201.78) and (373.77,246.41) .. (420,252.27) .. controls (441.72,255.03) and (453.72,257.02) .. (480,256.38) .. controls (484.55,256.26) and (489.23,253.18) .. (489.23,250.23) ;
\draw  [color={rgb, 255:red, 74; green, 144; blue, 226 }  ,draw opacity=1 ][fill={rgb, 255:red, 74; green, 144; blue, 226 }  ,fill opacity=1 ] (270.5,176.25) .. controls (270.5,173.35) and (272.85,171) .. (275.75,171) .. controls (278.65,171) and (281,173.35) .. (281,176.25) .. controls (281,179.15) and (278.65,181.5) .. (275.75,181.5) .. controls (272.85,181.5) and (270.5,179.15) .. (270.5,176.25) -- cycle ;
\draw  [color={rgb, 255:red, 74; green, 144; blue, 226 }  ,draw opacity=1 ][line width=1] [line join = round][line cap = round] (276,190.5) .. controls (276,194.45) and (280.02,197.53) .. (282,200.5) .. controls (287.59,208.88) and (292.82,217.96) .. (296,227.5) .. controls (301.25,243.24) and (289.61,255.96) .. (292,271.5) .. controls (294.56,288.16) and (314.11,287.8) .. (326,289.5) .. controls (328.71,289.89) and (335.65,294.5) .. (339,294.5) ;
\draw  [color={rgb, 255:red, 74; green, 144; blue, 226 }  ,draw opacity=1 ][line width=1] [line join = round][line cap = round] (270,269.5) .. controls (265.85,265.35) and (265.07,252.7) .. (263,246.5) .. controls (259.1,234.8) and (249.05,223.87) .. (244,212.5) .. controls (242.24,208.53) and (236.39,198.32) .. (238,193.5) .. controls (238.83,191.02) and (240.69,187.67) .. (244,187.5) .. controls (250,187.18) and (262,192.51) .. (262,186.5) ;
\draw  [color={rgb, 255:red, 74; green, 144; blue, 226 }  ,draw opacity=1 ][line width=1] [line join = round][line cap = round] (195.42,124.01) .. controls (191.51,124.01) and (189.58,131.52) .. (188.28,133.92) .. controls (181.15,147.13) and (175.6,158.93) .. (183.52,173.58) .. controls (188.06,181.99) and (219.62,177.02) .. (223.97,176.89) .. controls (236.66,176.52) and (249.36,176.89) .. (262.05,176.89) ;
\draw  [color={rgb, 255:red, 74; green, 144; blue, 226 }  ,draw opacity=1 ][line width=1] [line join = round][line cap = round] (266.78,170.52) .. controls (249.7,170.52) and (224.44,159.2) .. (217.97,146.43) .. controls (215.33,141.21) and (215.34,123.8) .. (221.46,117.76) .. controls (225.91,113.37) and (233.71,102.94) .. (241.22,101.7) .. controls (249.96,100.26) and (257.02,101.24) .. (264.46,99.41) .. controls (271.88,97.58) and (285.12,90.23) .. (291.19,90.23) ;
\draw  [color={rgb, 255:red, 74; green, 144; blue, 226 }  ,draw opacity=1 ][line width=1] [line join = round][line cap = round] (274.46,164.39) .. controls (271.68,164.39) and (269.02,151.75) .. (268.23,148.5) .. controls (264.85,134.58) and (265.68,114.55) .. (276.54,108.17) .. controls (302.49,92.89) and (335.13,99.61) .. (364.81,99.61) ;
\draw  [color={rgb, 255:red, 74; green, 144; blue, 226 }  ,draw opacity=1 ][line width=1] [line join = round][line cap = round] (283.93,162.34) .. controls (282.49,131.09) and (311.2,127.19) .. (328.43,123.41) .. controls (333.93,122.21) and (342.57,117.31) .. (349.07,116.79) .. controls (383.02,114.07) and (421.03,112.69) .. (447.3,143.46) .. controls (448.2,144.51) and (448.14,147.88) .. (448.66,149.07) .. controls (453.07,159.03) and (463.33,174.28) .. (470.03,182.12) ;
\draw  [color={rgb, 255:red, 74; green, 144; blue, 226 }  ,draw opacity=1 ][line width=1] [line join = round][line cap = round] (298.53,166.62) .. controls (312.36,166.62) and (332.24,155.66) .. (346.88,153.87) .. controls (378.26,150.05) and (404.08,164.56) .. (429.2,176.81) .. controls (435.1,179.69) and (439.31,188.03) .. (444.88,192.11) .. controls (452.74,197.85) and (462.49,199.93) .. (471.01,202.3) .. controls (473.12,202.89) and (481.47,206.05) .. (481.47,206.13) ;
\draw  [color={rgb, 255:red, 74; green, 144; blue, 226 }  ,draw opacity=1 ][line width=1] [line join = round][line cap = round] (323.14,175.83) .. controls (336,175.83) and (349.82,187.59) .. (361.23,192.3) .. controls (379.44,199.81) and (406.17,203.82) .. (427.57,207.5) .. controls (445.7,210.61) and (465.11,220.17) .. (482.86,220.17) ;

\end{tikzpicture}

        \caption{A singular foliation consists of a partitioning of a smooth manifold into dovetailing layers (``leaves''), much like a mille-feuille cake, except that the dimensions of the leaves can vary.}\label{fig:foliation}
    \end{figure}
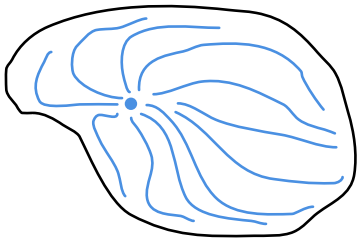

    \paragraph{Singular foliation associated to a Lie algebroid.}
    The partition of the moduli space \(M_0\) into equivalence classes is well known to mathematicians as a \emph{singular foliation}, which is a partition of a smooth manifold into dovetailing layers (or ``leaves'') of submanifolds, possibly of different dimensions, as shown in \cref{fig:foliation}.\footnote{`Singular' refers to the fact that different leaves may have different dimensions; the singular foliation is called ``regular'' when all leaves have the same dimension. So, confusingly, a regular foliation is a special case of a singular foliation.} More abstractly, a singular foliation \(F\) on a manifold \(M_0\) is given by a space of vector fields \(X^i\partial/\partial x^i\in F\) that are closed under the Lie bracket and multiplication by smooth functions and obey a technical finiteness condition; these vector fields are to be thought of as the tangent vector fields to the leaves.
    The above construction corresponds to the singular foliation associated to a Lie algebroid \(E\to M_0\), given by the collection of vector fields of the form
    \begin{equation}
        X^i(v) = s^a(v)\rho_a^i(v)
    \end{equation}
    at \(v\in M_0\), where \(s^a\) is an arbitrary section of the vector bundle \(E\) and where \(\rho_a^i\) is the anchor map of the Lie algebroid. The image of the anchor map \(\rho\) corresponds to the gauged directions, so the partition into S-type deformation equivalence classes is precisely the singular foliation associated to the gauge Lie algebroid.
    
    The dimension of each leaf determines the amount of spontaneous symmetry breaking: In a dimension \(k\), codimension \(d\) leaf, \(k\) scalar fields have been eaten by the gauge fields, leaving only \(d\) light scalars. The unbroken gauge symmetry is given by the \emph{isotropy algebra bundle} restricted to the leaf.
    Thus, the singular foliation associated to the effective gauge Lie algebroid controls the qualitative behavior of spontaneous symmetry breaking across the moduli space of vacua.
    
\begin{figure}\centering

\tikzset{every picture/.style={line width=0.75pt}}
\begin{tikzpicture}[x=0.75pt,y=0.75pt,yscale=-1,xscale=1]
\draw  [color={rgb, 255:red, 74; green, 144; blue, 226 }  ,draw opacity=1 ] (145.21,119.85) -- (228.8,77.67) -- (243.49,168.35) -- (159.91,210.54) -- cycle ;
\draw    (125.02,216.77) .. controls (127.02,180.77) and (188.35,168.1) .. (194.35,144.1) ;
\draw    (234.13,111) .. controls (248.13,93.67) and (238.8,77.67) .. (289.47,85) ;
\draw [color={rgb, 255:red, 74; green, 144; blue, 226 }  ,draw opacity=1 ]   (198.8,151.67) .. controls (232.25,190.21) and (234.71,138.48) .. (219.74,115.69) ;
\draw [shift={(218.8,114.33)}, rotate = 53.97] [color={rgb, 255:red, 74; green, 144; blue, 226 }  ,draw opacity=1 ][line width=0.75]    (10.93,-3.29) .. controls (6.95,-1.4) and (3.31,-0.3) .. (0,0) .. controls (3.31,0.3) and (6.95,1.4) .. (10.93,3.29)   ;
\draw [color={rgb, 255:red, 74; green, 144; blue, 226 }  ,draw opacity=1 ]   (197.47,139) .. controls (239.93,131.81) and (196.6,107.98) .. (169.13,120.83) ;
\draw [shift={(167.47,121.67)}, rotate = 331.78] [color={rgb, 255:red, 74; green, 144; blue, 226 }  ,draw opacity=1 ][line width=0.75]    (10.93,-3.29) .. controls (6.95,-1.4) and (3.31,-0.3) .. (0,0) .. controls (3.31,0.3) and (6.95,1.4) .. (10.93,3.29)   ;
\draw [color={rgb, 255:red, 74; green, 144; blue, 226 }  ,draw opacity=1 ]   (184.13,141) .. controls (152.28,129.95) and (152.75,167.7) .. (171.95,178.86) ;
\draw [shift={(173.47,179.67)}, rotate = 205.82] [color={rgb, 255:red, 74; green, 144; blue, 226 }  ,draw opacity=1 ][line width=0.75]    (10.93,-3.29) .. controls (6.95,-1.4) and (3.31,-0.3) .. (0,0) .. controls (3.31,0.3) and (6.95,1.4) .. (10.93,3.29)   ;
\draw [color={rgb, 255:red, 74; green, 144; blue, 226 }  ,draw opacity=1 ]   (190.8,154.33) .. controls (174.63,181.49) and (192.97,182.94) .. (213.55,180.56) ;
\draw [shift={(215.47,180.33)}, rotate = 172.87] [color={rgb, 255:red, 74; green, 144; blue, 226 }  ,draw opacity=1 ][line width=0.75]    (10.93,-3.29) .. controls (6.95,-1.4) and (3.31,-0.3) .. (0,0) .. controls (3.31,0.3) and (6.95,1.4) .. (10.93,3.29)   ;
\draw (101.33,70.33) node [anchor=north west][inner sep=0.75pt]   [align=left] {\begin{minipage}[lt]{50.37pt}\setlength\topsep{0pt}
\textcolor[rgb]{0.29,0.56,0.89}{transverse}
\begin{flushright}
\textcolor[rgb]{0.29,0.56,0.89}{model }
\end{flushright}

\end{minipage}};
\draw (263.33,88.33) node [anchor=north west][inner sep=0.75pt]   [align=left] {leaf};

\end{tikzpicture}
\caption{The transverse model to a leaf captures the pattern of vacuum deformations near a leaf.}\label{fig:transverse-model}
    \end{figure}
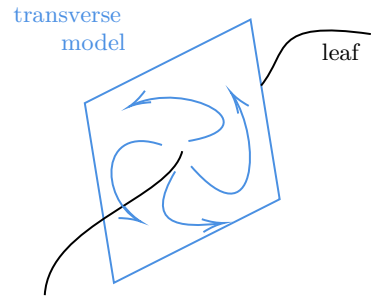

    \paragraph{Transverse model and possible behavior of spontaneous symmetry breaking.}
    Given a leaf \(L\), the directions parallel to it are by definition S-type deformations staying in the same phase while the directions transverse to it are G-type or S-type deformations to other phases. However, the structure of the transverse directions is in general far more complicated. In the directions transverse to \(L\), one has the \emph{transverse model} \(T\) \cite[Def.\ 1.26]{2401.05966}, which captures the behavior of G-type and S-type deformations near it (\cref{fig:transverse-model}): For a codimension \(d\) leaf, the transverse model $T$ is a singular foliation on \(\mathbb R^d\) such that, for a sufficiently small local patch \(U\subset L\), the foliation is isomorphic to \(U\times T\) in a neighborhood of \(U\); by definition, the vector fields along \(T\) vanish at \(0\).
    
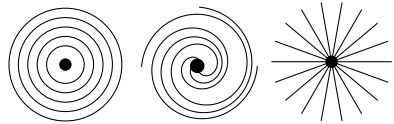
\begin{figure}\centering
     \begin{tabular}{ccc}
     \begin{tikzpicture}[scale=0.5]
\filldraw[fill=white](0,0) circle (1.5);
\filldraw[fill=white](0,0) circle (1.25);
\filldraw[fill=white](0,0) circle (1);
\filldraw[fill=white](0,0) circle (0.75);
\filldraw[fill=white](0,0) circle (0.5);
\filldraw[fill=black](0,0) circle (0.15);
\end{tikzpicture}
    &
 \begin{tikzpicture}[scale=0.4]
 \node[rotate=90] at (0.3,0.2) {
    \begin{tikzpicture}[scale=0.4, node distance = 0cm, auto](0,0)\spiral[black][0,0][0:2:2] ;
    \end{tikzpicture}};
 \node[rotate=180] at (-0.1,0.2) {
    \begin{tikzpicture}[scale=0.4, node distance = 0cm, auto](0,0)\spiral[black][0,0][0:2:2] ;
    \end{tikzpicture}};
    \spiral[black][0,0][0:2:2];
  \filldraw[color=black!100, fill=black!100, very thick](0,0) circle (0.1875);
\end{tikzpicture}
 &
\begin{tikzpicture}[scale=0.8]
    \draw (0,0) -- (0:1cm);
    \draw (0,0) -- (20:1cm);
    \draw (0,0) -- (40:1cm);
    \draw (0,0) -- (60:1cm);
    \draw (0,0) -- (80:1cm);
    \draw (0,0) -- (100:1cm);
    \draw (0,0) -- (120:1cm);
    \draw (0,0) -- (140:1cm);
    \draw (0,0) -- (160:1cm);
    \draw (0,0) -- (180:1cm);
    \draw (0,0) -- (-20:1cm);
    \draw (0,0) -- (-40:1cm);
    \draw (0,0) -- (-60:1cm);
    \draw (0,0) -- (-80:1cm);
    \draw (0,0) -- (-100:1cm);
    \draw (0,0) -- (-120:1cm);
    \draw (0,0) -- (-140:1cm);
    \draw (0,0) -- (-160:1cm);
    \filldraw[fill=black](0,0) circle (0.09375);
\end{tikzpicture}
\end{tabular}
        \caption{
            Some possible two-dimensional transverse models around the leaf \(\bullet\):
            S-deformations for the vector field \(a(x\partial_y-y\partial_x)+b(x^2+y^2)(x\partial_x+y\partial_y)\) for \((a,b)=(1,0)\), \((a,b)=(1,1)\), and \((a,b)=(0,1)\).
            In the first case, S-type deformations can never reach the leaf at the origin. In the second and third cases, S-type deformations can reach the leaf with a phase transition at the end.
        }\label{fig:foo}
    \end{figure}

    The behavior of the transverse foliation can be varied. In some cases, S-type deformations can never reach the leaf \(L\), as in the first figure of \cref{fig:foo}. In some other cases, S-type deformations can reach the leaf (possibly at asymptotically infinite length), as in the second and third cases of \cref{fig:foo}, with a phase transition at the end into a phase with less broken gauge symmetry (because of lower dimension of leaf). The space of such trajectories can be entwined in topologically complicated ways.
    
    \section{Classification of vacuum deformations near a given vacuum orbit}
    
    A natural question now is the following: Given a submanifold \(L\subset M_0\) inside the vacuum moduli space which we wish to regard as a vacuum orbit (leaf), what are the possible vacuum deformations (transverse models) near this vacuum orbit? It may seem that one would be able to conclude little without detailed knowledge of the gauge algebroid. However, a recent theorem \cite{2401.05966} shows that in fact there are powerful constraints on the possible patterns of vacuum deformations near a vacuum orbit and one can explicitly compute whether a given vacuum-deformation pattern is possible for a given topology of the vacuum orbit.
    
    Let \(L\) be a vacuum orbit of codimension \(d\) with transverse model \(T\). Given this, one has the \emph{inner automorphism group} \(\operatorname{Inn}(T)\) consisting of invertible analytic maps \(\phi\colon\mathbb R^d\to\mathbb R^d\) that are flows of (i.e.~arise from integrating) vector fields that belong to \(T\). This Lie group can be infinite-dimensional, so it is convenient to quotient it by the subgroup \(\operatorname{Inn}_{\ge2}(T)\) corresponding to those coordinate transformations generated by vector fields that vanish quadratically near the origin. Then the quotient \(G(T)=\operatorname{Inn}(T)/\operatorname{Inn}_{\ge2}(T)\) is guaranteed to be a finite-dimensional Lie group. We will call \(G(T)\) the \emph{leading-order flow group} of the transverse model \(T\).
    
    Now, one key statement \cite[Cor.\ 3.5]{2401.05966} implies that, if \(L\) does not have non-contractible paths (i.e.\ is simply connected), then singular foliations admitting $L$ as a leaf and $T$ as transverse model are in one-to-one correspondence to principal \(G(T)\)-bundles. More concretely, every principal \(G(T)\)-bundle over \(M_0\) can be lifted uniquely to a principal \(\operatorname{Inn}(T)\)-bundle, and the normal bundle of \(L\) inside \(M_0\) is an associated bundle of the \(\operatorname{Inn}(T)\)-bundle.\footnote{The \(\operatorname{Inn}(T)\)-action may not be linear, so this is not always an associated vector bundle.} That is, given a principal \(G(T)\)-bundle on the vacuum orbit, we can reconstruct the pattern of vacuum deformations near the vacuum orbit.
    
    One may worry that we have classified ``too much,'' that is, we may have singular foliations that do not correspond to Lie algebroids that can appear in physics. In fact every singular foliation is realizable as the phases of a Lie algebroid.\footnote{Namely: given a transverse model \(T\), the Atiyah algebroid of the \(\operatorname{Inn}(T)\)-bundle canonically acts on the normal bundle, and the associated action algebroid realizes a singular foliation with transverse model \(T\); see \cite[App.~A]{2401.05966}.} The resulting Lie algebroid always admits locally a connection and \(\zeta\) satisfying \eqref{eq:adjustment_cond} \cite{Fischer:2024vak}. One caveat is that in fact the Lie algebroids may be infinite-dimensional (corresponding to infinitely many fields); alternatively, one can instead have a physical theory with finitely many fields but with higher-order \(p\)-form fields \cite{Laurent-Gengoux:1806.00475}.
    
    The classification and statements above are local in the sense that it only holds in a suitable neighborhood around the leaf \(L\).
    Furthermore, we have restricted our attention to the case when \(L\) lacks non-contractible paths (i.e.~\(L\) is simply connected). If there are non-contractible paths, the classification becomes richer; see \cite[Thm.\ 2.16]{2401.05966}. Furthermore, the result above is formal in the sense that it ignores analytical issues relating to convergence of formal series. For a fuller mathematical treatment, see \cite{2401.05966}.
    
    \section{Examples}
    We start with two extreme examples (assuming that the leaf $L$ is simply connected). Consider the case where the transverse model \(T\) is given by the vector field that is zero everywhere, i.e.~when all deformations are G-type and there are no S-type deformations. In that case, there are no vector fields to take flows of, so \(G(T)\) and \(\operatorname{Inn}(T)\) are the trivial group; all leaves are of the same dimension (namely, zero), and hence there are no phase transitions. In that case, for a simply-connected leaf \(L\), the normal bundle must be a trivial bundle.\footnote{More generally, whenever the \(\operatorname{Inn}(T)\)-bundle admits a flat connection, then the \(\operatorname{Inn}(T)\)-bundle is trivial and the normal bundle of the simply connected leaf \(L\) is therefore also trivial.}
    
    On the other hand, consider the case where the transverse model \(T\) for a leaf \(L\) of dimension \(k\) and codimension \(d\) consists of all vector fields that vanish at the origin so that there are two phases (one corresponding to \(L\), with \(d\) Goldstone bosons and \(k\) massive vector bosons; the other corresponding to the neighborhood of \(L\), with no Goldstone bosons and \(d+k\) massive vector bosons). Then the group \(\operatorname{Inn}(T)\) generated by the flows consists of arbitrary origin-preserving diffeomorphisms whose differentials at the origin have positive determinants; the leading-order flow group is therefore \(G(T)=\operatorname{GL}_+(d)\), the group of \(d\times d\) real matrices with positive determinants.   
		Since arbitrary orientable\footnote{Every fiber bundle with fiber $\mathbb{R}^d$ over a simply connected base is orientable.} bundles with fiber $\mathbb{R}^d$ can be written as associated bundles of \(\operatorname{Inn}(T)\)-bundles,
		there are no constraints on the topology of the normal bundle in this case; all such bundles are possible.
    
    The generic case is intermediate between these two. Consider, for example, the family of two-dimensional transverse models \(T_{a,b}\) generated by the vector fields
    \begin{equation}
        a(x\partial_y-y\partial_x)+b(x^2+y^2)(x\partial_x+y\partial_y)
    \end{equation}
    for some specific values of \(a\) and \(b\), as shown in \cref{fig:foo}. The corresponding leading-order flow groups are
    \begin{equation}
        G(T_{a,b})=\begin{cases}
            \operatorname U(1) & \text{if \(a \ne 0 = b\)}, \\
            \mathbb R & \text{if \(a\ne 0\ne b\)}, \\
            1 & \text{if \(a=0\)}.
        \end{cases}
    \end{equation}
    Suppose for instance that \(L=S^2\) is the two-dimensional sphere that is the base of the local Calabi--Yau manifold \(\mathrm T^*S^2\), a toy version of the famous deformed conifold \(\mathrm T^*S^3\) that appears in string theory compactifications. This would mean that the normal bundle to \(L\) would be the cotangent bundle, which is topologically nontrivial in this case. So, for this \(L\), the transverse model \(T_{a,b}\) is possible for \(a\ne0=b\) (since there are nontrivial \(\operatorname U(1)\)-bundles over \(\mathbb{CP}^1\)) but not in the other cases (since there are no nontrivial \(\mathbb R\)-bundles or \(1\)-bundles over \(\mathbb{CP}^1\)).
    
    \

    \noindent\emph{Data Management.} No additional research data beyond the data presented and cited in this work are needed to validate the research findings in this work.
    
    \
    
        \noindent \emph{Acknowledgments.}  S.-R.F.\ thanks Camille Laurent-Gengoux for helpful comments. S.-R.F.\ was supported by a postdoctoral fellowship at the National Center for Theoretical Sciences, Taipei. M.J.F.\ was supported by the STFC PhD studentship ST/W507489/1. H.K.\ thanks Luigi Alfonsi and Leron Borsten for helpful comments. H.K.\ was partially supported by the Leverhulme Research Project Grant RPG-2021-092. C.S.\ was partially supported by the Leverhulme Research Project Grant RPG-2018-329.
    \bibliographystyle{../latexeu2}
    \bibliography{../bigone,extra}
\end{document}